\documentclass[11pt, fleqn]{article}

\def\be{\begin{equation}}
\def\ee{\end{equation}}

\def\cou{4\pi\epsilon_0\,}

\usepackage{amsmath}
\usepackage{amsfonts, pstcol,times,color,graphicx,graphics,pstricks,pst-node,pst-coil,pst-grad, pst-plot, pst-3d, pst-fill, multido, pst-poly, url,bm, pst-math, pst-xkey, pstricks-add, pst-eucl, hyperref, pst-func}
\newpsobject{showgrid}{psgrid}{subgriddiv=1,griddots=10,gridlabels=6pt}

\begin{document}

\title{A note on the electrostatic energy of two point charges }
\author{ A C Tort\footnote{email: tort@if.ufrj.br.}\\
Instituto de F\'{\i}sica
\\
Universidade Federal do Rio de Janeiro\\
Caixa Postal 68.528; CEP 21941-972 Rio de Janeiro, Brazil}
\maketitle
\begin{abstract}
The electrostatic field energy due to two fixed point-like charges  shows some peculiar features concerning the distribution in space of the field energy density of the system. Here  we discuss the evaluation of the field energy and the  mathematical details that lead to those peculiar and non-intuitive physical features.
\end{abstract}
\pagestyle{myheadings}
\markright{ a c tort  2013}

\newpage
\section{Introduction}
The electrostatic energy of two point charges is given by the simple expression

\begin{equation}
U_{12}=\frac{1}{4\pi\epsilon_0} \,\frac{q_1\,q_2}{R} , \label{U12}
\end{equation}
where $R$ is distance between the charges the values of which are $q_1$ and $q_2$.  If the charges have the same algebrical sign the electrostatic energy is positive but if the algebrical signs are not equal then electorstatic energy is negative. Equation (\ref{U12}) is interpreted as a potential energy due to the spatial configuration of the charges. If constraints forces are removed this energy will be transformed into kinetic energy of the charges. On the other hand, from the field point of view the  energy of an electrostatic system is given by the  expression

\begin{equation}
U = \frac{1}{2\epsilon_0}\, \iiint  \,\|\mathbf{E} (P) \|^2\, dV \geq 0 , 
\end{equation}
where $\mathbf{E}(P)$ is the total field of the electrostatic system at a point $P$ of the space, that is

\be \mathbf{E} (P)  = \mathbf{E}_1 (P) + \mathbf{E}_2 (P).  \ee
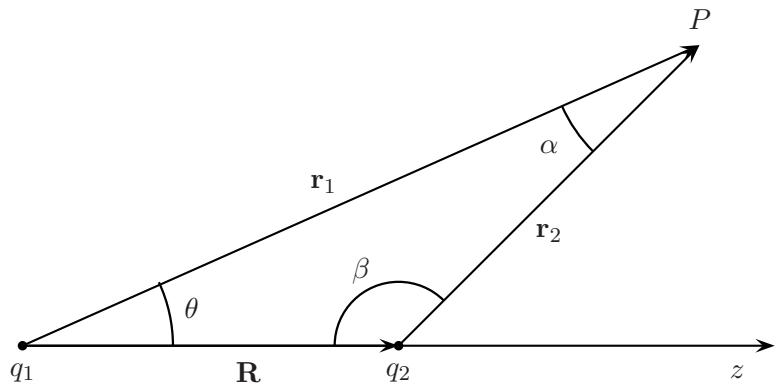
\begin{figure}[!h]
\begin{center}
\begin{pspicture}(-6,-1)(6,5)
\psset{arrowsize=0.1 3}
\psdot(-5, 0)
\psdot(0,0)
\psline{->}(-5,0)(5,0)
\psline{->}(0,0)(4,4)
\psline{->}(-5,0)(4,4)
\psline{->}(-5,0)(0,0)
\psarc[linewidth=0.30mm]{-}(-5,0){2}{0}{25}
\psarc[linewidth=0.30mm]{-}(0,0){0.85}{45}{180}
\psarc[linewidth=0.30mm]{-}(4,4){2}{204}{225}
\rput(4.5, -0.35){$z$}
\rput(-5, -0.35){$q_1$}
\rput(0, -0.35){$q_2$}
\rput(-1,2.15){$\mathbf{r}_1$}
\rput(2, 1.5){$\mathbf{r}_2$}
\rput(-2.75, 0.5){$\theta$}
\rput(-0.5,1){$\beta$}
\rput(2,2.65){$\alpha$}
\rput(4, 4.35){$P$}
\rput(-2, -0.35){$\mathbf{R}$}
\end{pspicture}
\caption{Geometry for the evaluation of the interaction energy of two point charges.}
\label{vectors}
\end{center}
\end{figure}

\newpage
\noindent The field energy can be split into three separate contributions

\begin{equation} 
\frac{\epsilon_0}{2}\, \iiint \, \mathbf{E}^2(P)\, dV = \frac{\epsilon_0}{2}\, \iiint \, \mathbf{E}_1^2(P)\, dV  +  \frac{\epsilon_0}{2}\, \iiint \, \mathbf{E}_2^2(P)\, dV + \epsilon_0 \, \iiint \, \mathbf{E}_1(P)\cdot\mathbf{E_2}(P)\,dV.  \label{U}
 \end{equation}
The first two terms on RHS of (\ref{U}) can be interpreted as the (divergent) self-energies of the point charges and in a classical context simply hid under the rug. The crossed term is interpreted as the interaction energy of the two charges and must reproduce (\ref{U12}), that is 

\be  \epsilon_0 \, \iiint \, \mathbf{E}_1(P)\cdot\mathbf{E_2}(P)\,dV = \frac{1}{4\pi\epsilon_0} \,\frac{q_1\,q_2}{R} , \ee
but this must proved by explicit evaluation of integral.  
\vskip 10pt
The first time the author heard of this problem was when he was reading the first edition of reference  \cite{Berkeley} where it was presented as an advanced problem in a special chapter at the end of the book. The challenge was to prove that the potential energy point of view and the field energy one were not incompatible. Recently, in a paper on the role of field energy in introductory physics courses Hilborn \cite{Hilborn} commented on this and some peculiar features concerning the distribution in space of the field energy of the system. To understand these features consider for simplicity two equal positive charges  then (1) there is a spherical region centered at one of the charges that does not contribute to the total interaction energy; (2) this region can be divided into two subregions that contribute with algebraically opposite energies and the amount of negative energy is very small when compared with the total energy,  and finally $90\%$ of the field energy lies in a limited part of the space, see the appendix in  \cite{Hilborn}.  Here the present author will try to show to the interested reader the details of those peculiar and non-intuitive aspects by performing explicitly the calculations. 
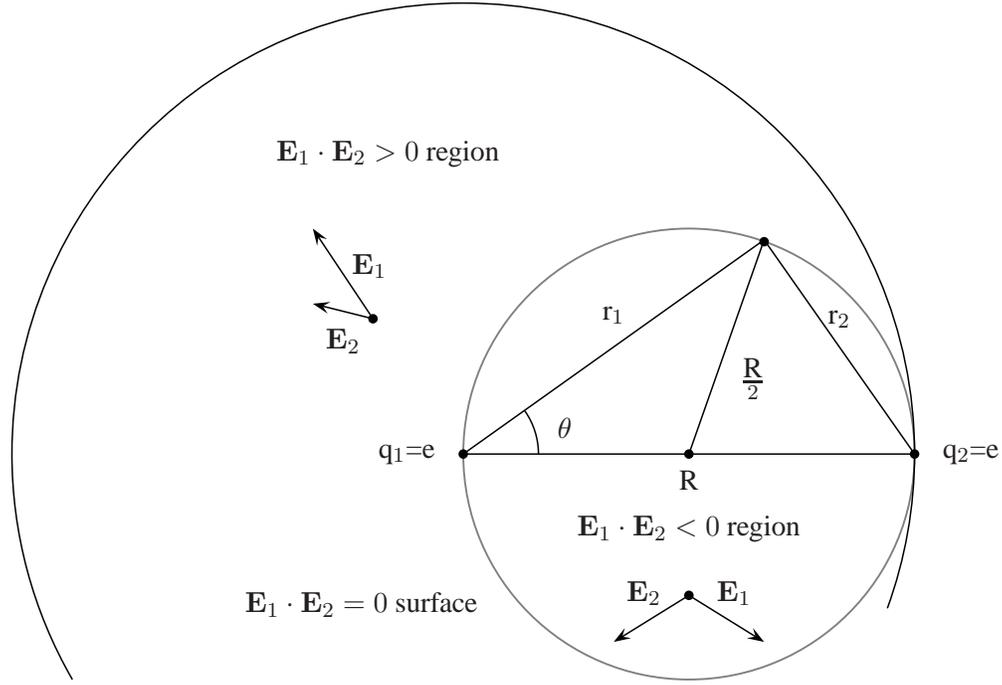
\begin{figure}[!h]
\begin{center}
\begin{pspicture}(-9,-3)(5,6)
\psset{arrowsize=0.1 3}

\psline[linewidth=0.20mm]{-}(-3,0)(3,0)
\psline[linewidth=0.20mm]{-}(-3,0)(1,2.85)
\psline[linewidth=0.20mm]{-}(3,0)(1,2.85)
\psline[linewidth=0.20mm]{-}(0,0)(1,2.85)
\psarc[linewidth=0.20mm]{-}(-3,0){1}{0}{35}
\rput(-1.65,0.35){$\theta$}
\rput(-3.75, 0){q$_1$=e}
\rput(3.75, 0){q$_2$=e}
\pscircle[linewidth=0.25mm, linecolor=gray](0,0){3}
\psarc[linewidth=0.20mm]{-}(-3,0){6}{-20}{210}
\rput(-1,1.85){r$_1$}
\rput(2, 1.80){r$_2$}
\rput(0.85,1){$\frac{\mbox{R}}{2}$}

\rput(0,-0.35){R}
\rput(-4,4){{$\mathbf E_1\cdot\mathbf E_2 >  0$ region}}
\rput(0,-1){{$\mathbf E_1\cdot\mathbf E_2 <  0$ region}}
\rput(-4.25, 2.5){$\mathbf E_1$}
\rput(-4.60, 1.5){$\mathbf E_2$}

\rput(0.60, -1.85){$\mathbf E_1$}
\rput(-0.60, -1.85){$\mathbf E_2$}

\psline[linewidth=0.20mm]{->}(-4.20,1.80)(-5,3)
\psline[linewidth=0.20mm]{->}(-4.20,1.80)(-5,2)
\psdot(-4.20, 1.80)

\psline[linewidth=0.20mm]{->}(0,-1.88)(-1,-2.5)
\psline[linewidth=0.20mm]{->}(0,-1.88)(1,-2.5)
\psdot(0,-1.88)
\rput(-4.35,-2){$\mathbf E_1\cdot\mathbf E_2 = 0$ surface}
\psdot(-3, 0)
\psdot(3,0)
\psdot(1,2.825)
\psdot(0,0)
\end{pspicture}
\caption{The distribution of the electrostatic energy density for two equal positive point charges.}
\label{sphere1}
\end{center}
\end{figure}

\section{Evaluation of the electrostatic energy}
It is convenient to use spherical coordinates with the origin at one of the charges and the polar $z$-axis along the line that passes through both, see Figure \ref{vectors}. Let us set $q_1=q_2=e$. The crossed term reads

\be  \mathbf{E}_1\cdot\mathbf{E_2} = \frac{e^2 }{(4\pi\epsilon_0)^2 r_{1}^2 r_{2}^2 }\, \mathbf{e}_{r_1} \cdot \mathbf{ e}_{r_2} . \ee
Now we introduce, see Figure \ref{vectors}

\be \mathbf{r}_1=\mathbf{R} + \mathbf{r}_2 \ee
hence

\be r_2^2=r_1^2 + R^2 -2r_1 R\,\cos\, \theta ,  \label{r2} \ee
and, see Figure \ref{vectors}
\be \hat{e}_{r_1}\cdot \hat{e}_{r_2}= \cos\,\alpha . \ee
Therefore

\be \mathbf{E}_1\cdot\mathbf{E_2} = \frac{e^2}{(4\pi\epsilon_0)^2} \,\frac{\cos\,\alpha }{r_1^2 \left(r_1^2 + R^2 -2 r_1 R \, \cos\, \theta  \right)} . \ee 
To relate $\alpha$ and $\theta$  we apply the sine law to the triangle in Figure \ref{vectors}

\be \frac{R}{\sin\,\alpha} =\frac{r_2}{\sin\,\theta}  .\ee
Combining this relation with the fundamental trignometric identity $\cos^2\,\alpha +\sin^2\,\alpha=1$, we find after some simple manipulations

\be \cos\,\alpha =  \sqrt{ 1- \frac{R^2}{r_2^2} \left(1- \cos^2\,\theta \right)}. \ee
Now we take (\ref{r2}) into the equation above and after some simplifications we get

\be 1- \frac{R^2}{r_2^2} \left(1- \cos^2\,\theta \right) =  \frac{\left(r_1-R \cos\,\theta\right)^2}{r_1^2 + R^2 -2r_1 R\,\cos\, \theta } . \ee
It follows that

\be \cos\,\alpha =   \frac{r_1-R \cos\,\theta}{\sqrt{r_1^2 + R^2 -2r_1 R\,\cos\, \theta }} , \ee
and

\be  \mathbf{E}_1\cdot\mathbf{E_2}= \frac{e^2 }{(4\pi\epsilon_0)^2}\, \frac{r_1-R \cos\,\theta}{r_1^2\left(r_1^2 + R^2 -2r_1 R\,\cos\, \theta\right)^{3/2} } . \label{dotP} \ee
Therefore we must now compute

\be U_{12}=\epsilon_0 \, \times \, \frac{e^2 }{(4\pi\epsilon_0)^2} \, \int _0^\infty \,r_1^2 \, dr_1 \, \int_{\Omega} \, \frac{r_1-R \cos\,\theta}{r_1^2\left(r_1^2 + R^2 -2r_1 R\,\cos\, \theta\right)^{3/2} }  \, d\Omega \ee
where $d\Omega = \sin\,\theta\,d\theta \, d\phi$.  The integration over the azimuthal angle is trivial and yields a factor $2\pi$, and upon introducing the variable $\xi=\cos\,\theta$ we have

\be  U_{12}= \frac{e^2}{8\pi\epsilon_0} \ \int_0^\infty \, f(r_1) \, dr_1 \label{U12b},\ee
where we have defined

\be f(r_1) = \int_{-1}^{+1} \, \frac{r_1-R\,\xi}{\left(r_1^2 + R^2 -2r_1 R\, \xi\right)^{3/2} }\, d\xi .\ee
This integral can evaluated straightforwardly and because $R>0$ , $r_1 > 0 $, we can write the result as

\be f(r_1)= \frac{|R-r_1|\, (R+r_1) -(R+r_1) \, (R-r_1) }{|R-r_1|\,(R+r_1)\,r_1^2}  . \label{fr1value}\ee
To proceeed we must consider separately two cases. 
\vskip 10pt
\noindent\underline{Case $r_1 <  R$}. In  this case it is easily seen that

\be f(r_1)= 0 . \ee
This means that inside of an imaginary sphere of radius equal to $R$, the interaction energy of the two charges is zero. 
\vskip 10pt
\noindent\underline{Case $r_1 >  R$}. In this case, Eq. (\ref{fr1value}) yields

\be f(r_1)= \frac{2}{r_1^2} . \ee
Taking this result into Eq. (\ref{U12b}) we obtain the expected result

\be U_{12} = \frac{e^2}{\cou} \, \int_R^\infty \, \frac{dr_1}{r_1^2} = \frac{e^2}{\cou R}. \ee
Notice that if we set the upper limit equal to $10R$, then a simple calculation shows that

\be U^{\,\prime }_{12} = \frac{e^2}{\cou} \, \int_R^{10R} \, \frac{dr_1}{r_1^2} = \frac{9}{10} \frac{e^2}{\cou R}, \ee

\noindent that is, as stated in \cite{Hilborn}, $90\%$ of energy is contained between two spheres, one of radius $R$ and the other one of radius $10R$. If the charges are not identical, all we have to do is replace $e^2$ by $q_1q_2$. 
\section{The energy density distribution}
Though the final result is the one we expected the way it was obtained reveals some details that are somewhat surprising, to wit, the part of the field energy that corresponds to the interaction energy of the two point charges comes from the region $r > R$. The region $r < R$ makes no contribution at all. This conclusion agrees with reference \cite{Hilborn}. How can this physically be? 
\vskip 10pt
To answer  this question we  must first realize that the electrostatic interaction energy density of the system is essentially given by the dot product of the fields. In the case of two positive identical charges it is not difficult to see that the angle between the fields, let us denote it by $\alpha$ as before, is obtuse near the charges and acute far away from them, see Figure \ref{sphere1}. In fact, the negative contributions comes from a spherical region of radius equal to $R/2$ centered at the midpoint between the two charges, see Figure \ref{sphere1}. The positive contribution comes from the rest.   The volume of the spherical region is smaller than the volume of the rest, but the fields are more intense near the charges  than far away from them.  Therefore, we conclude that in the end there is a cancellation between the corresponding contributions and this is reason why the entire region $r < R$ makes no contribution at all to the final result. The electrostatic field energy of this system comes from the region $R < r < \infty$.   

\vskip 10pt
Inside the region $r < R$ there is a surface that separates the negative energy density region from the positve one. On this surface the fields are perpendicular to each other and the energy density is null. This can be seen from Eq. (\ref{dotP}).  The dot product between the fields is zero if and only if

\be  r_1=R\,\cos\,\theta , \ee   
but if we inspect Figure \ref{sphere1} this relation is a consequence of Thales' theorem that states that any triangle inscribed in a semicircle is a right triangle which is the case of the triangle formed by the three segments of line whose lengths are $r_1$, $r_2$, and $R$. It follows easily that  on the spherical surface of radius equal to $R/2$ the fields are perpendicular to each other and consequently the interaction energy is zero. 
\newpage
\section{The ratio between negative and positive energy}
\begin{figure}[!t]
\begin{center}
\begin{pspicture}(-6,-2)(6,5)
\psset{arrowsize=0.1 3}

\psarc[linewidth=0.20mm]{-}(-4,0){1}{0}{23.0}
\psarc[linewidth=0.20mm]{-}(0,0){1}{0}{63.4}
\psarc[linewidth=0.20mm]{-}(0,0){4}{-10}{190}
\rput(-2.65,0.25){$\theta$}
\rput(1.25,0.65){$\theta^{\,\prime}$}
\rput(-4.75, 0){q$_1$=e}
\rput(4.75, 0){q$_2$=e}
\rput(-1,1.65){$\mathbf r_1$}
\rput(2.5, 1.50){$\mathbf r_2$}
\rput(0,1){$\mathbf X$}

%
\rput(-2, -0.35){$\mathbf x_1$}
\rput(2, -0.35){$\mathbf x_2$}

\psline[linewidth=0.20mm]{-}(0,0)(1.05,2.10)
\psline[linewidth=0.20mm]{->}(-4,0)(4,0)
\psline[linewidth=0.20mm]{<-}(-4,0)(0,0)
%
\psline[linewidth=0.20mm]{->}(4,0)(0,2.85)
\psline[linewidth=0.20mm]{->}(-4,0)(2,2.5)
\psdot(1.05,2.10)
\psline[linewidth=0.20mm]{->}(4,0)(1.05,2.10)
\psline[linewidth=0.20mm]{->}(-4,0)(1.05,2.10)
\rput(-2.35,4.5){{$\mathbf E_1\cdot\mathbf E_2 = 0$ surface}}
\psdot(-4, 0)
\psdot(4,0)
\psdot(0,0)
\end{pspicture}
\caption{Geometry for the evaluation of the negative energy density contribution.}
\label{vectors3}
\end{center}
\end{figure}
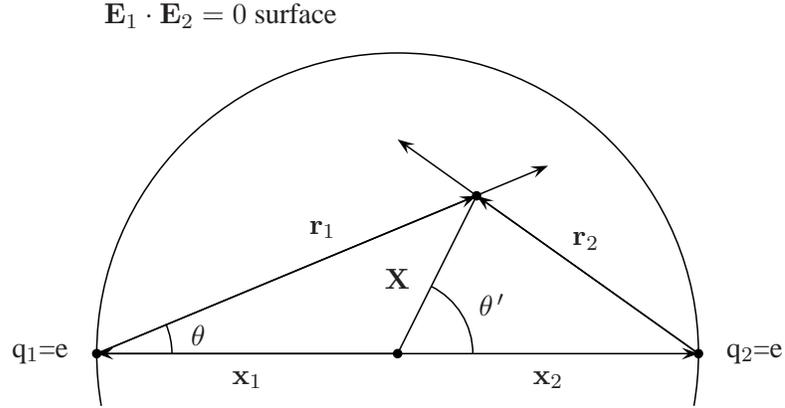

Let us now evaluate the ratio between negative and positive field interaction energy. The hard part is the evaluation of the contribution of the negative energy. In order to perform this calculation some geometrical transformations will have to be made. Consider Figure \ref{vectors3}.  In order to shift the origin of the coordinate system to the midpoint between the charges we introduce the position vectors of the charges with respect to the midpoint, $\mathbf x_1$ and $\mathbf x_2$, such that $\mathbf x_1 + \mathbf x_2=0$ and $\|\mathbf x_1\|=\|\mathbf x_2\|=R/2$. We introduce also the position vector $\mathbf X$ of an arbitrary point $P$ with respect to the midpoint.  Notice that the magnitude $\| \mathbf X\|=X$ of this vector lies in the interval  $0 \leq  \mathbf X  \leq R/2$.  The new polar angle is $\theta^{\,\prime}$ and the following vector relations are easily seen to hold
\be \mathbf r_1 =\mathbf x_2 +\mathbf X ;  \hskip 2cm \mathbf r_2=-\mathbf x_2 + \mathbf X . \ee
\noindent The interaction energy as before depends on the dot product

\be  \mathbf{E}_1\cdot\mathbf{E_2} = \frac{e^2}{(4\pi\epsilon_0)^2 r_{1}^2 r_{2}^2 }\, \mathbf{e}_{r_1} \cdot \mathbf{ e}_{r_2}= \frac{e^2 }{(4\pi\epsilon_0)^2 r_{1}^2 r_{2}^2 }\, \frac{\mathbf r_1}{r_1} \cdot \frac{\mathbf r_2}{r_2} . \ee
From the vector relations above it follows that

\be r_1^2 = \frac{R^2}{4} +X^2 + R X\, \cos\,\theta^{\,\prime}  , \ee
and 
\be r_2^2 = \frac{R^2}{4} +X^2 - R X\, \cos\,\theta^{\,\prime}  . \ee
We also have

\be \frac{\mathbf r_1}{r_1} \cdot \frac{\mathbf r_2}{r_2} = \frac{X^2-\displaystyle{\frac{R^2}{4}}}{r_1 r_2} .\ee
Defining the adimensional variable 

\be u:=\frac{X}{R} ,  \hskip 1cm 0\leq u \leq 1, \ee
the field energy content of this region $A_{12}$ can be written as

\be
A_{12}= U_{12}\, \int_0^{1/2}\, du\, u^2 \left(u^2 -\frac{1}{4}\right)  \int_0^1 \frac{d\xi }{\left( \frac{1}{4} +u^2+ u\,\xi \right)^{3/2}\,\left( \frac{1}{4} +u^2- u\,\xi \right)^{3/2} } ,
\ee
where $\xi : = \cos\,\theta^{\,\prime}$.  The integral in $\xi$ is

\be  \int_0^1 \frac{d\xi }{\left( \frac{1}{4} +u^2+ u\,\xi \right)^{3/2}\,\left( \frac{1}{4} +u^2- u\,\xi \right)^{3/2} } =  \frac{64}{ \|2u-1\|\,(2u+1) \left( 16 u^4 +8u^2 +1 \right)} , \ee
and after performing the integral in $u$ we obtain

\be A_{12}=- U_{12} \, \frac{\pi -2}{4} .\ee
In order to get zero energy inside the spherical region of radius $R$ centered at one of the charges we must have an equal amount of a positive contribution $B_{12}$

\be B_{12} = + U_{12} \, \frac{\pi -2}{4} .\ee
Therefore the ratio of the negative energy to the positive energy is

\be \frac{A_{12}}{U_{12} + B_{12}} = -\left(\frac{\pi -2}{\pi+2}\right) \approx - 0.2220 . \ee
This means that the negative energy content is considerable less than the positive energy one. 

\section{Final remarks}
To conclude let us call the reader's attention to two points. The first one is that if the point charges have opposite algebraic signs then as before there will still be a sphere centered at one of the charges of radius equal to their separation inside of which the total content of energy is zero, but the energy contained in the smaller sphere of radius $R/2$ will be positive and the rest of the energy will be negative. This can be easily seen by sketching the dot product $\mathbf E_1\cdot\mathbf E_2$. 

\vskip 10pt
The second one is that all calculations done for the configuration considered here -- \emph{mutatis muntandis} -- apply to the corresponding gravitational case. The content of energy stored in the gravitational field is given by

\be U = - \frac{1}{8\pi G}\, \iiint \, \mathbf g^{\,2}(P) \,dV ,\ee
where $\mathbf g(P)$ is the resultant field at a point $P$. For a gravitational configuration similar to the electrostatic one considered here $\mathbf g(P)=\mathbf g_1(P)+\mathbf g_2(P)$, and depending on the model we choose for the mass distribution we will not need to hid infinities due to self-energies under the rug. The same can be said about charge distributions, though here we dealt with point charges.  

\end{document}